\documentclass[conference]{IEEEtran}

\IEEEoverridecommandlockouts

\usepackage[english]{babel} 

\usepackage{cite}

\ifCLASSINFOpdf
   \usepackage[pdftex]{graphicx}
\else
   \usepackage[dvips]{graphicx}
   \DeclareGraphicsExtensions{.eps}
\fi

\usepackage{grffile}
\usepackage{epstopdf}

\usepackage{subfig}

\usepackage[cmex10]{amsmath}
\usepackage{color}
\usepackage{array}
\begin{document}

\title{On the Stability of Asynchronous Random Access Schemes}

\author{\IEEEauthorblockN{Alessio Meloni
\thanks{A. Meloni gratefully acknowledges Sardinia Regional Government for the financial support of his PhD scholarship (P.O.R. Sardegna F.S.E. 2007-2013 - Axis IV Human Resources, Objective l.3, Line of Activity l.3.1.).}and Maurizio Murroni\thanks{\copyright 2013 IEEE. The IEEE copyright notice applies. DOI: 10.1109/IWCMC.2013.6583667}}
\IEEEauthorblockA{DIEE - Department of Electrical and Electronic Engineering\\
University of Cagliari\\
Piazza D'Armi, 09123 Cagliari, Italy\\
Email: \{alessio.meloni\}\{murroni\}@diee.unica.it}
}


\maketitle

\begin{abstract}
Slotted Aloha-based Random Access (RA) techniques have recently regained attention in light of the use of Interference Cancellation (IC) as a mean to exploit diversity created through the transmission of multiple burst copies per packet content (CRDSA). Subsequently, the same concept has been extended to pure ALOHA-based techniques in order to boost the performance also in case of asynchronous RA schemes. In this paper, throughput as well as packet delay and related stability for asynchronous ALOHA techniques under geometrically distributed retransmissions are analyzed both in case of finite and infinite population size. Moreover, a comparison between pure ALOHA, its evolution (known as CRA) and CRDSA techniques is presented, in order to give a measure of the achievable gain that can be reached in a closed-loop scenario with respect to the previous state of the art.\\ 
\end{abstract}

\begin{IEEEkeywords}
Random Access, Aloha, Interference Cancellation, CRDSA, CRA, Stability, Equilibrium Contour, Packet Delay
\end{IEEEkeywords}

\IEEEpeerreviewmaketitle

\section{Introduction}

Despite its almost 40 years-long life since the original idea was published \cite{AbramsonALOHA}, Aloha and its successive evolutions such as Slotted Aloha (SA) \cite{RobertsALOHA} and Diversity Slotted Aloha (DSA) \cite{DiversityALOHA} have been always used in many random access application scenarios (such as initial terminal login in satellite communications), especially in case of long propagation delay and directive transmissions that do not allow carrier sensing and collision avoidance as for example in 802.11 DCF.
Basically all Aloha-based techniques have in common the capability to allow transmissions from a number of terminals in a multi-access channel without the need of coordination among them, even though this means that the possibility of collision between two or more different packets is present. 
Recently, these techniques and in particular synchronous access schemes (i.e. those in which the channel is divided into slots) have received new interest in light of a breakthrough idea that consists in introducing Interference Cancellation (IC) in DSA schemes. 
Differently from SA in which packets are sent just once (or once per communication feedback in case of a system using retransmissions), in DSA multiple burst copies are sent for the same packet. It has been demonstrated that the diversity created by multiple transmissions is beneficial in case of small channel load while achieves worse results from moderate channel loads on. 
The idea behind the use of IC in DSA is to further exploit the advantage of sending multiple copies by trying to restore also the content of colliding packets. This new access scheme is known as Contention Resolution Diversity Slotted Aloha (CRDSA) \cite{CRDSA1} and works as follows. In CRDSA, terminals transmit packets in a given frame (composed of a certain number of slots) by placing two packet's burst copies in two randomly chosen slot locations. Each burst copy contains a pointer identifying the slot position of its twin. At the receiver, if at least one burst copy of a given packet can be correctly decoded, its signal content is removed from all other involved slots thanks to IC. By iteratively repeating this procedure, it is possible to restore the content of those packets that had all their burst copies interfering, if at least one burst copy interfered with bursts belonging to correctly decoded packets that are thus eligible for IC. As a result, while original SA technique reaches a peak throughput $T\simeq 0.37 [pkt/slot]$, CRDSA boosts the performance up to $T\simeq 0.55 [pkt/slot]$. 

Further studies have regarded the optimization of the number of copies per packet (namely burst degree) to be sent. In particular  \cite{CRDSA2} and \cite{CRDSA3} deal with the use of more than two copies per packet, demonstrating by means of simulation that the results can be beneficial in terms of maximum achievable throughput and/or in terms of Packet Loss Ratio depending on the chosen burst degree. For example, the use of 3 copies per packet yields to a throughput peak $T\simeq 0.68 [pkt/slot]$ while using 5 copies can lower the Packet Loss Ratio down to $PLR=1\cdot 10^{-6}$ for load values up to $G=0.6 [pkt/slot]$.
Afterwards, the same idea has been extended to the case of Irregular Burst Degree, known as Irregular Repetition Slotted Aloha (IRSA)\cite{IRSA1} and renamed in the DVB-RCS2 Lower Layer Satellite Specification \cite{RCS2} as Variable Rate - Contention Resolution Diversity Slotted Aloha (VR-CRDSA). In this case the number of copies per each packet is chosen accordingly to a given burst degree probability distribution that is optimized via differential evolution, allowing to reach throughput values up to $T\simeq 0.8 [pkt/slot]$ in practical implementations. Last but not least, as with the birth of Slotted Aloha techniques a certain interest on the related stability in case of retransmissions came out \cite{STAB1} \cite{STAB2}, also the birth of CRDSA has given place to some works that analyze its stability in case of retransmissions and compare its performance with the one for SA  \cite{CRDSA_stab1} \cite{CRDSA_stab3}. CRDSA has currently been introduced as option for Random Access communication in DVB-RCS2 \cite{RCS2} and its use has been discussed in a quasi-real-time satellite mobile messaging systems \cite{CRDSA2}.

In a recent paper the same concept behind CRDSA has been applied to Pure Aloha giving birth to a technique called Contention Resolution Aloha (CRA) \cite{CRA}. For this reason, in this paper the analysis in terms of stability that has been carried out for CRDSA is extended to the case of asynchronous RA schemes. This analysis uses the same tools adopted for synchronous access schemes, with the necessary modifications needed in order to take into account the differences between the two techniques. The paper is organized as follows. In Section II an overview of the considered asynchronous access scheme is given. Section III presents the definition of stability as well as the model used for the measure of the stability when using retransmissions. Section IV deals with a model for the computation of the delay associated to received packets. Finally, in Section V a comparison both in terms of stability and in terms of delay is carried out between CRDSA, CRA and pure ALOHA. Section VI concludes the paper.   

\section{System Overview}\label{SO}

The scenario considered in this paper is a multi-access channel for satellite communications, in which a certain number of terminals communicate to a gateway (e.g. a ground station) via satellite. Differently from synchronous access schemes, in this case the channel is divided into frames but each frame is not subdivided into slots. When a terminal has a packet to transmit, it waits for the beginning of the next available frame in order to place $d$ copies of that packet. Let us call $t_0$ the beginning of a frame, $T_F$ the frame duration and $\tau$ a generic burst duration. The $d$ copies of a packet are placed with starting time within the interval [$t_0$;$t_0+T_F-\tau$], with uniformly distributed probability and so that burst copies belonging to the same packet are not overlapping.

\begin{figure}[h!]
\centering
\includegraphics [width=1 \columnwidth] {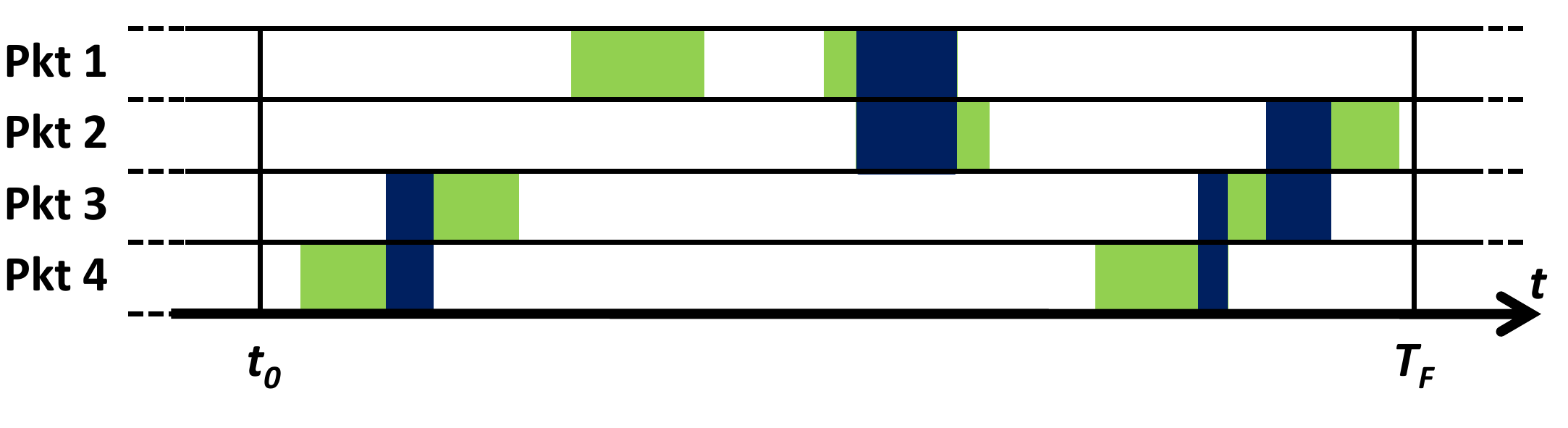}
\caption{\small{Example of a generic frame at the receiver for CRA. In light color (green) portions of the burst not overlapping. In dark color (blue) portions of the burst overlapping with other bursts}}
\label{CRA}
\end{figure}

At the receiver, each frame contains a certain number of bursts as depicted in Figure~\ref{CRA}. Any burst will have or not a certain degree of interference due to transmission time overlap with other bursts. Notice that differently from CRDSA, in which interference can only occur for the whole burst, in this case also partial interference can occur. In \cite{CRA} two cases are analyzed: the first case assumes that any overlap results in entire loss of the packet's burst; the second case considers the application of a strong FEC able to allow decoding if the amount of interference is limited. In any case, similarly to what happens in CRDSA, an iterative IC process is started at the receiver in order to remove bursts belonging to correctly decoded packets thanks to the knowledge of their location within the frame from the correctly decoded burst. 

Consider the assumption of ideal channel estimation and perfect Interference Cancellation. In the first case (i.e. where no FEC is used), at each iteration packet bursts are attempted to be decoded only if the burst is not overlapping with other bursts. In Figure~\ref{CRA}, $Packet\ 1$ has a copy that did not interfere during transmission, therefore it can be decoded and the interference of the other burst copy can be removed in order to recover the content of $Packet\ 2$. In the second case in which a strong FEC is applied, not only bursts without interference, but also those satisfying a certain threshold in terms of amount of interference power are decoded.
Let us define as in the original CRA paper \cite{CRA} the rate $R=R_C\cdot log_2M$ where $R_C$ is the coding rate and $M$ the modulation index. Moreover the normalized MAC channel load is defined as $G=\frac{N_{tx}\cdot \tau}{T_F}$ with $N_{tx}$ indicating the number of transmitted packets, while $T(G)=G [1-PLR(G)]$ represents the throughput in terms of portion of load successfully decoded. Notice that PLR (i.e. the Packet Loss Ratio) depends on the frame size $T_F$, the burst degree distribution (defined from \cite{IRSA1} as the probability of having a certain number of copies per packet through the following polynomial $\Lambda(x) = \sum_d \Lambda_d x^d$, where $\Lambda_d$ is the probability that a given packet will have burst degree $d$), the rate $R$, the maximum number of iterations for the decoding process $I_{max}$ and the $SNIR$.
In \cite{CRA} the decoding threshold has been approximated with the Shannon bound. As claimed by the authors, even though this threshold is quite optimistic, it can be considered valid for moderate to high SNIR with properly designed schemes and represent the ground base for the study of the performance with real codes. Setting $C=R=log_2(1+SNIR)$, the decoding threshold is $SNIR_{dec,dB}=10\cdot log_{10}(2^R-1)$. In order for a burst to be decoded, its $SNIR$ must be at least equal to $SNIR_{dec}$. The $SNIR$ of each burst can be computed as

\begin{figure}[t!]
\centering
\includegraphics [width=0.95 \columnwidth] {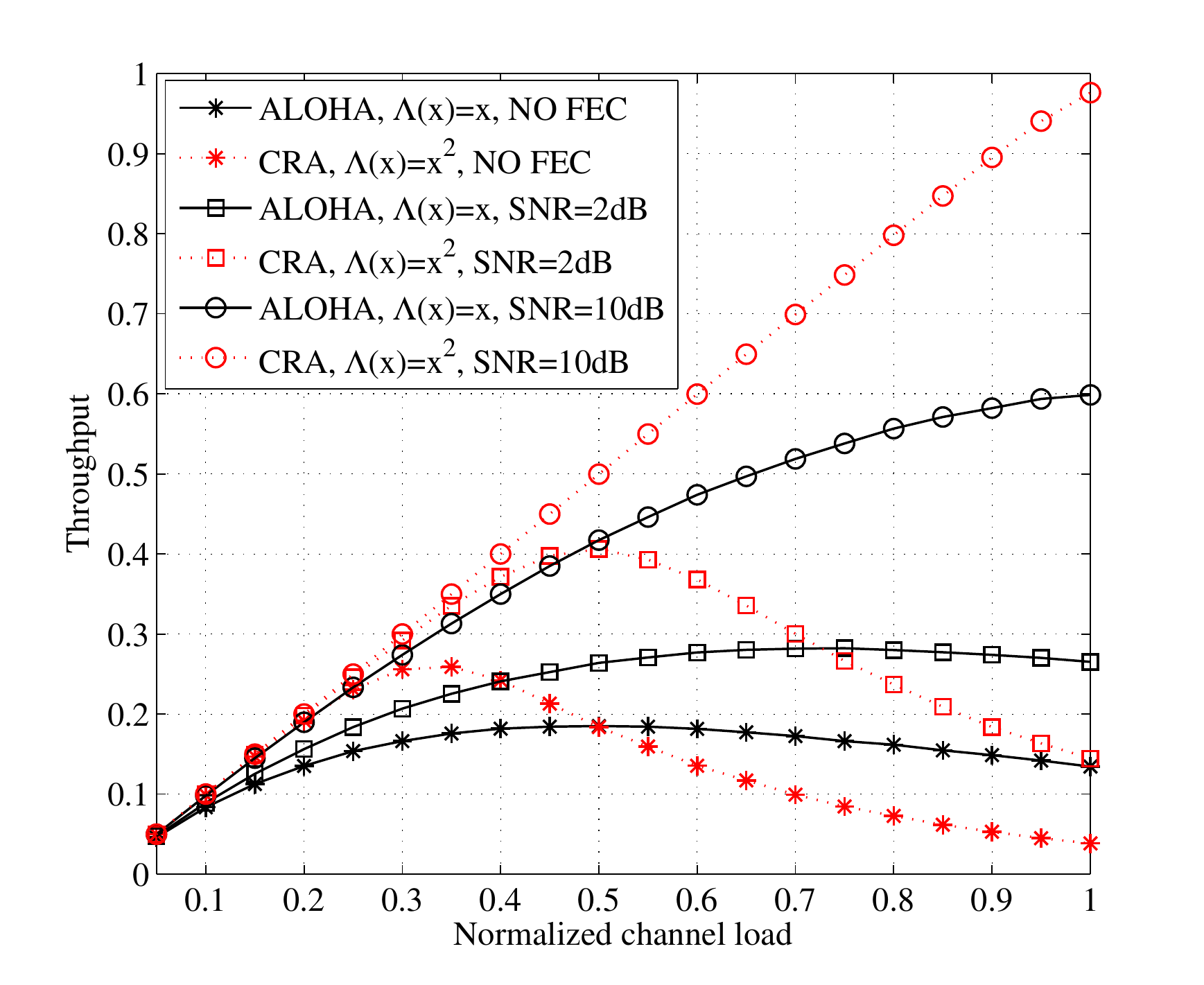}
\caption{\small{Open loop throughput results}}
\label{thrpop}
\end{figure}

\begin{equation}
SNIR=\frac{P}{x\cdot P+N}=\frac{SNR}{x\cdot SNR+1}
\end{equation}
where $x$ represents the degree of interference for a certain burst as a sum over all interference contributions expressed with a value between $0$ and $1$. For example, in case of no interference $x=0$, in case of $50\%$ overlapping with another burst $x=0.5$ and in case of $50\%$ interference with $n$ other bursts, $x=0.5\cdot n$. 
In Figure~\ref{thrpop} results for an open loop scenario (i.e. without retransmission of lost contents) are illustrated for ALOHA and the representative case of CRA with $\Lambda(x)=x^2$. The parameter values chosen for simulations are the same that will be used throughout the paper: $T_F=100\ ms$, $\tau=1\ ms$ for every packet, $M=4$ (QPSK), $R_C=1/2$, $I_{max}=50$ and a number of simulation rounds per channel load point equal to $10^4$.

\section{Stability}

Assume the case in which CRA has been chosen as random access scheme and a certain number of users $N_u$ (either finite or infinite) participate in the described scenario. We assume each user to be in one of two possible states: Thinking (T) or Backlogged (B). 

\begin{figure}[th!]
\centering
\includegraphics [width=0.95 \columnwidth] {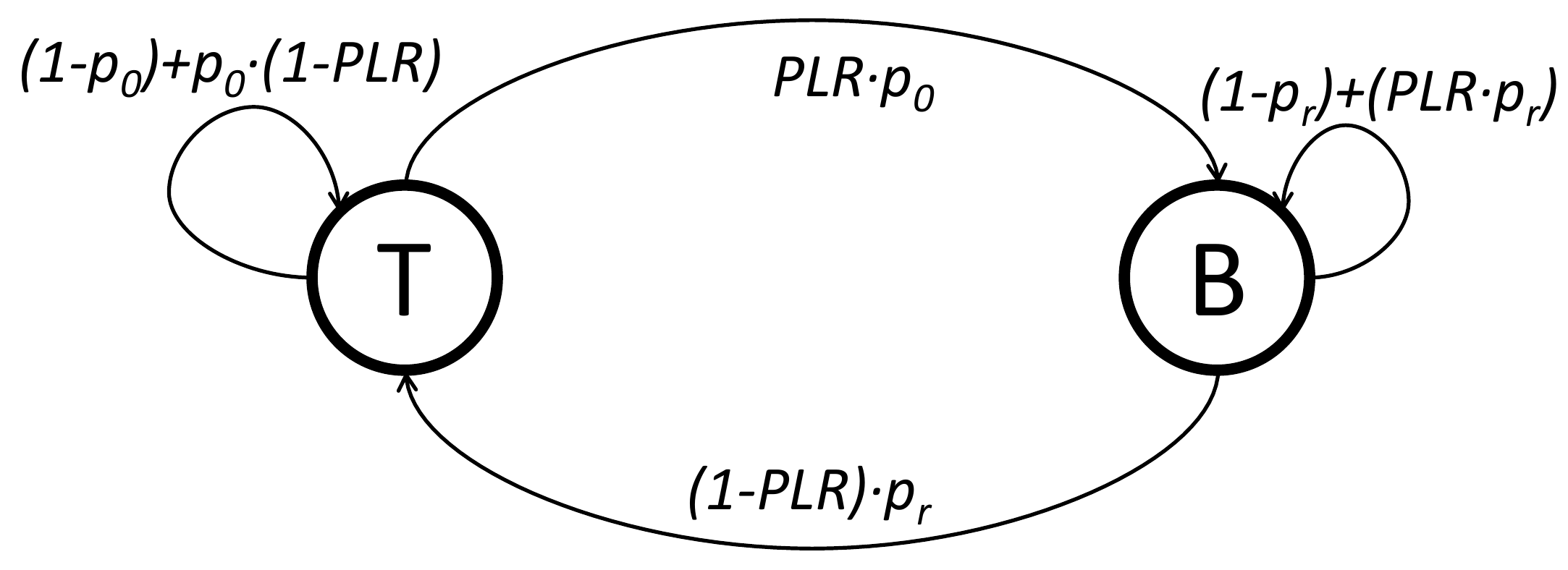}
\caption{\small{Markov Chain for user state}}
\label{user_state}
\end{figure}

Users in $T$ state send a packet at the beginning of the next frame with probability $p_0$. Assuming that users are acknowledged about  the outcome of the transmission at the end of the frame in which they transmitted, if the packet is correctly decoded the user stays in $T$ state. Therefore the probability of staying in Thinking state is equal to the probability that a user does not send any packet plus the probability that a user sends a packet that is correctly received at the first attempt. On the other hand, if a user is unsuccessful in its first attempt, it switches to backlogged state. In B state, a user attempts retransmission with probability $p_r$. In case the retransmission ends up successfully the user comes back to Thinking state at the end of the frame in which it retransmitted its packet while in case of no retransmission or unsuccessful retransmission, it stays in $B$ state.

Let us define $N_B^j$ as the number of backlogged packets at the end of frame $j$ and $N_{TOT}$ as the total number of users, so that
\begin{equation}
G_B^j=\frac{N_B^{(j-1)}\cdot \tau\cdot p_r}{T_F}
\end{equation}
is the expected channel load in frame $j$ due to retransmissions and
\begin{equation}\label{CHL}
G_T^j=\frac{(N_{TOT}-N_B^{(j-1)})\cdot \tau\cdot p_0}{T_F}
\end{equation}
is the expected channel load of frame $j$ due to new transmissions. Finally $G^j=G_T^j+G_B^j$ is the expected total channel load of frame $j$.

The aim of the definitions above is to find the \textit{equilibrium contour} in a plane having as axis the number of backlogged users and the channel load due to new transmissions. As a matter of fact, \textit{equilibrium contour} is defined as the locus of points for which the expected channel load due to new transmissions is equal to the expected throughput, so that the communication can be considered as stable and the total expected channel load $G^j$ is equal frame after frame.
The expected number of new transmissions at the equilibrium can be defined as
\begin{equation}\label{gt}
 G_T=T(G)=G\ [1-PLR(G)]
\end{equation}

In stability conditions, also the number of backlogged users remains the same frame after frame. Therefore
\begin{equation}\label{nb1}
N_B=N_B (1-p_r) + \frac{G\cdot T_F}{\tau} PLR(G)
\end{equation}
from which
\begin{equation}\label{nb2}
N_B=\frac{G\cdot PLR(G)\cdot T_F}{\tau\cdot p_r}
\end{equation}
Equations \eqref{gt} and \eqref{nb2} describe the \textit{equilibrium contour}. This contour, together with the expected channel load due to new transmissions in Equation~\ref{CHL} (known as channel load line) gives a model for the computation of the stability. 

\begin{figure}[tbh!]

\subfloat [Stable channel] {\label{stable} \includegraphics [ scale = 0.22 ]{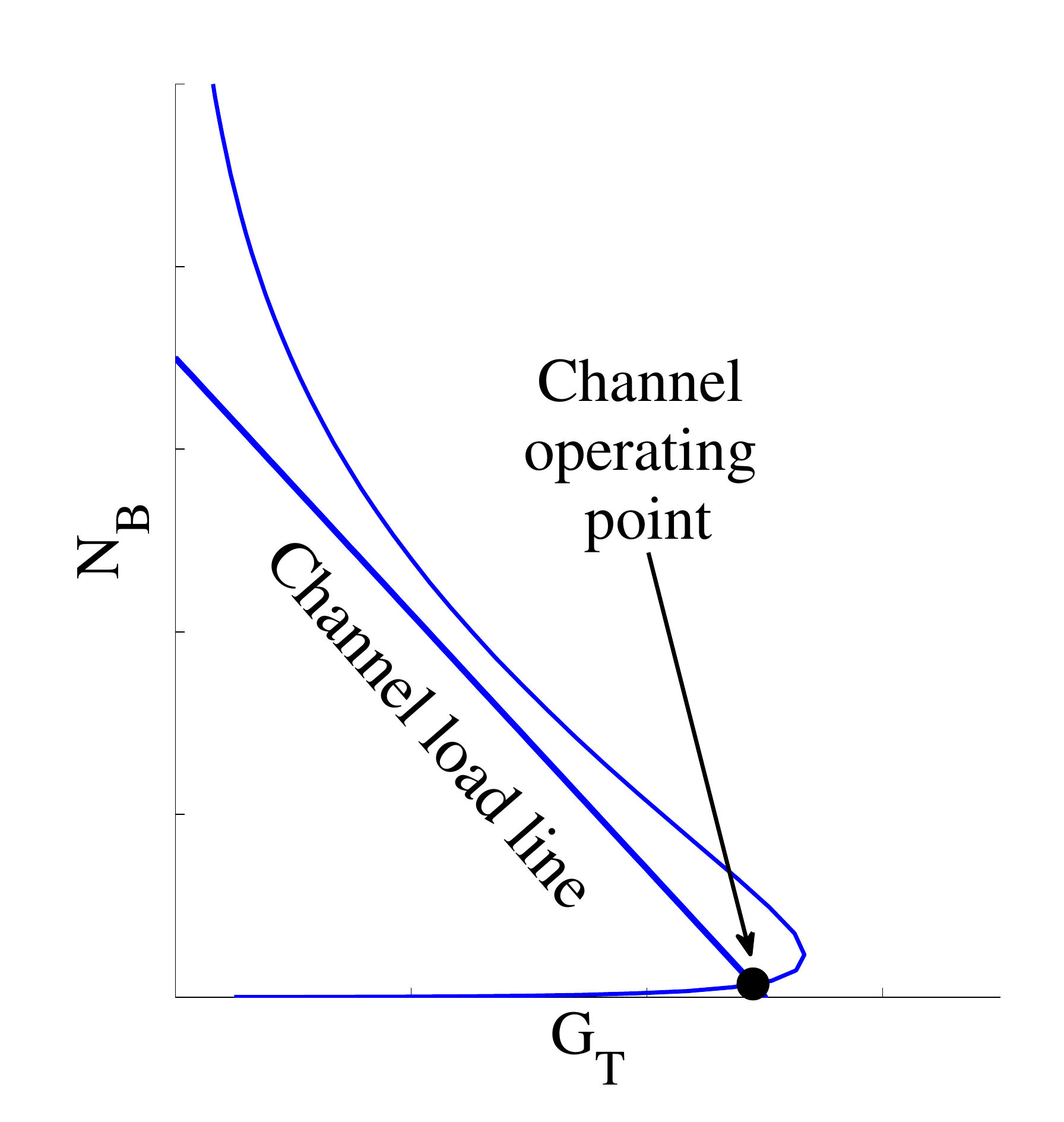}
\label{st}
} \qquad
\subfloat [Unstable channel (finite M)] {\label{unstableFin} \includegraphics [ scale = 0.22 ]{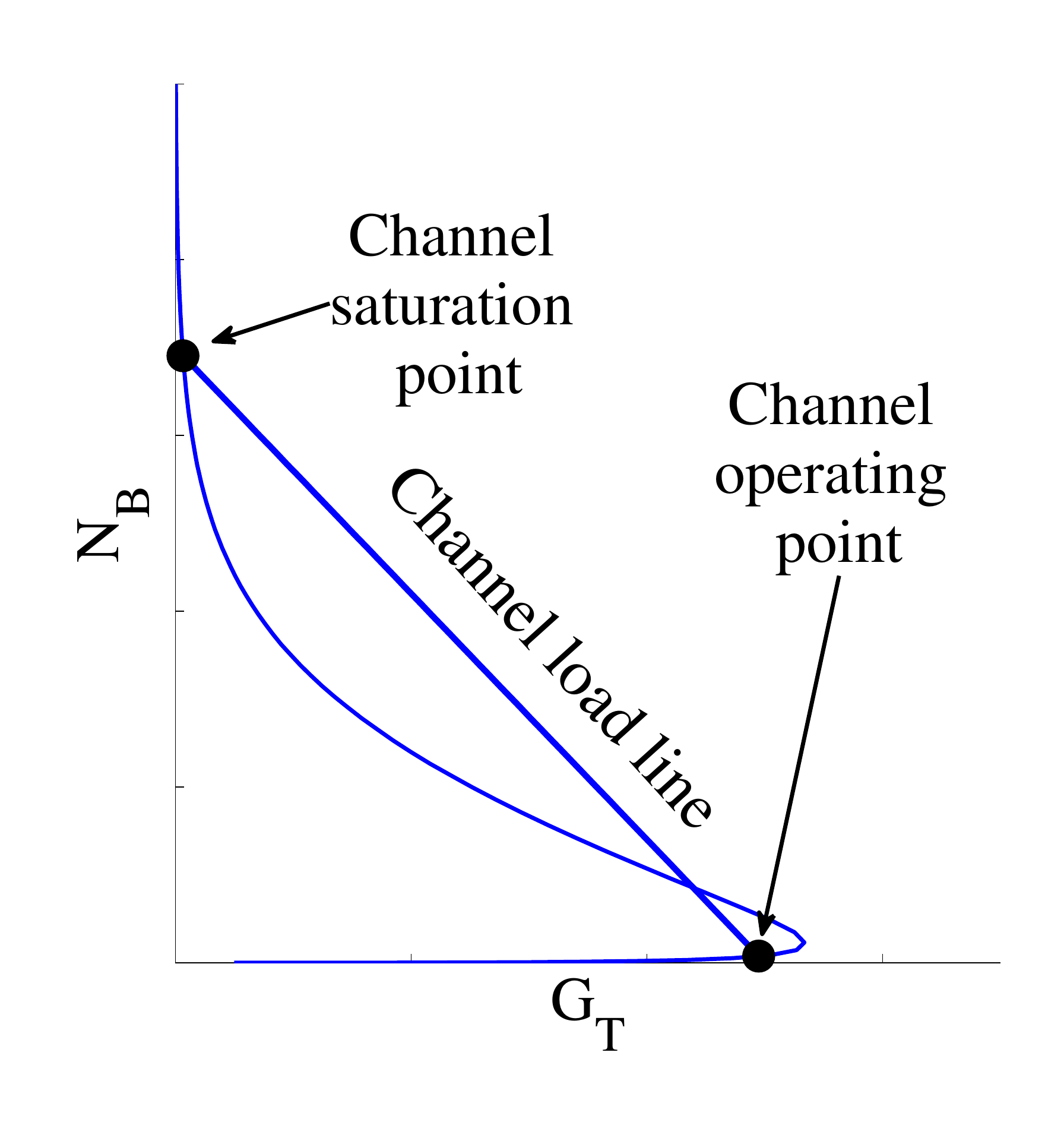}
\label{unstFin}
} \qquad

\subfloat [Unstable channel (infinite M)] {\label{unstableInf} \includegraphics [ scale = 0.22 ]{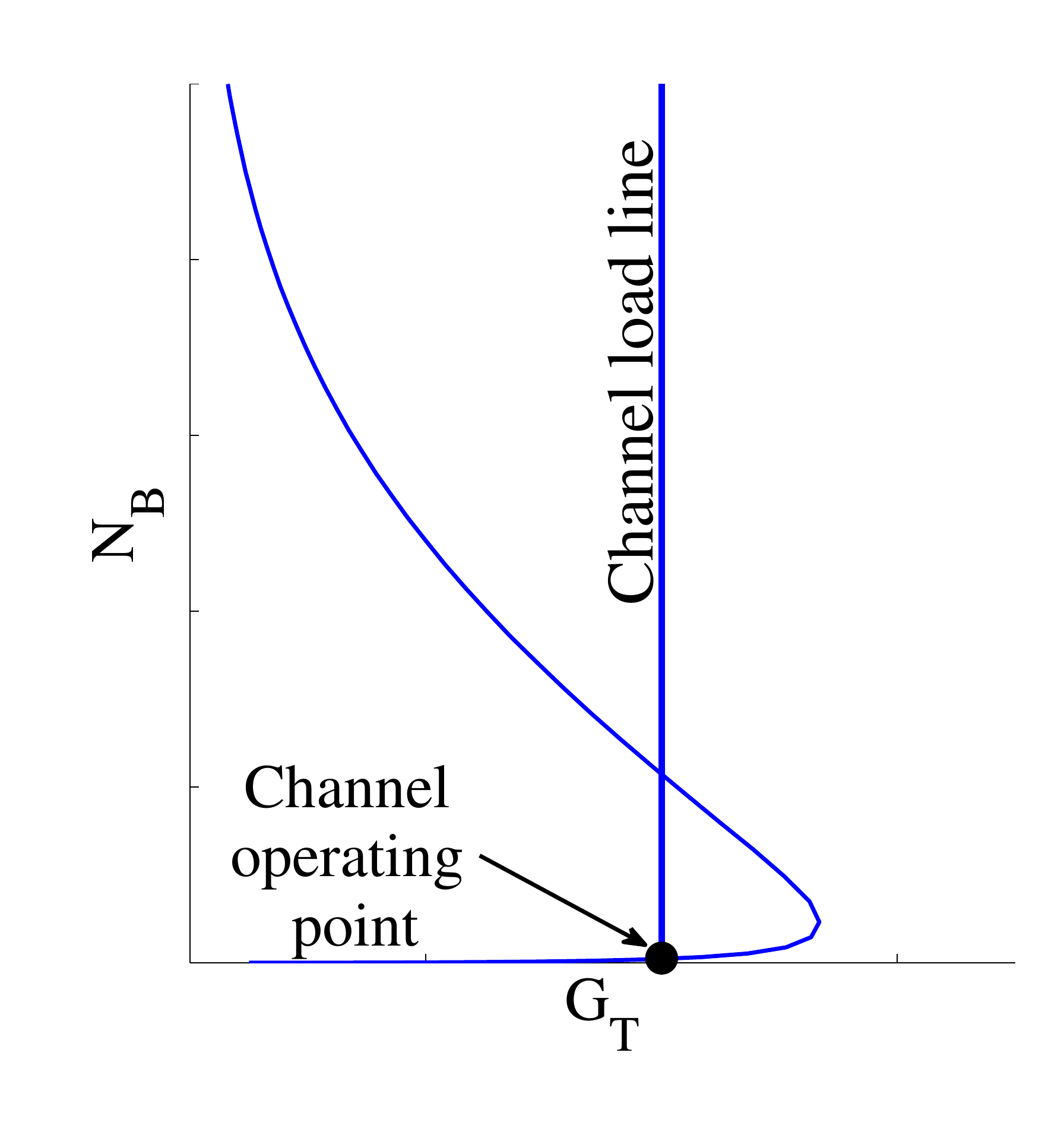}
\label{unstInf}
} \qquad
\subfloat [Overloaded channel] {\label{overL} \includegraphics [ scale = 0.22 ]{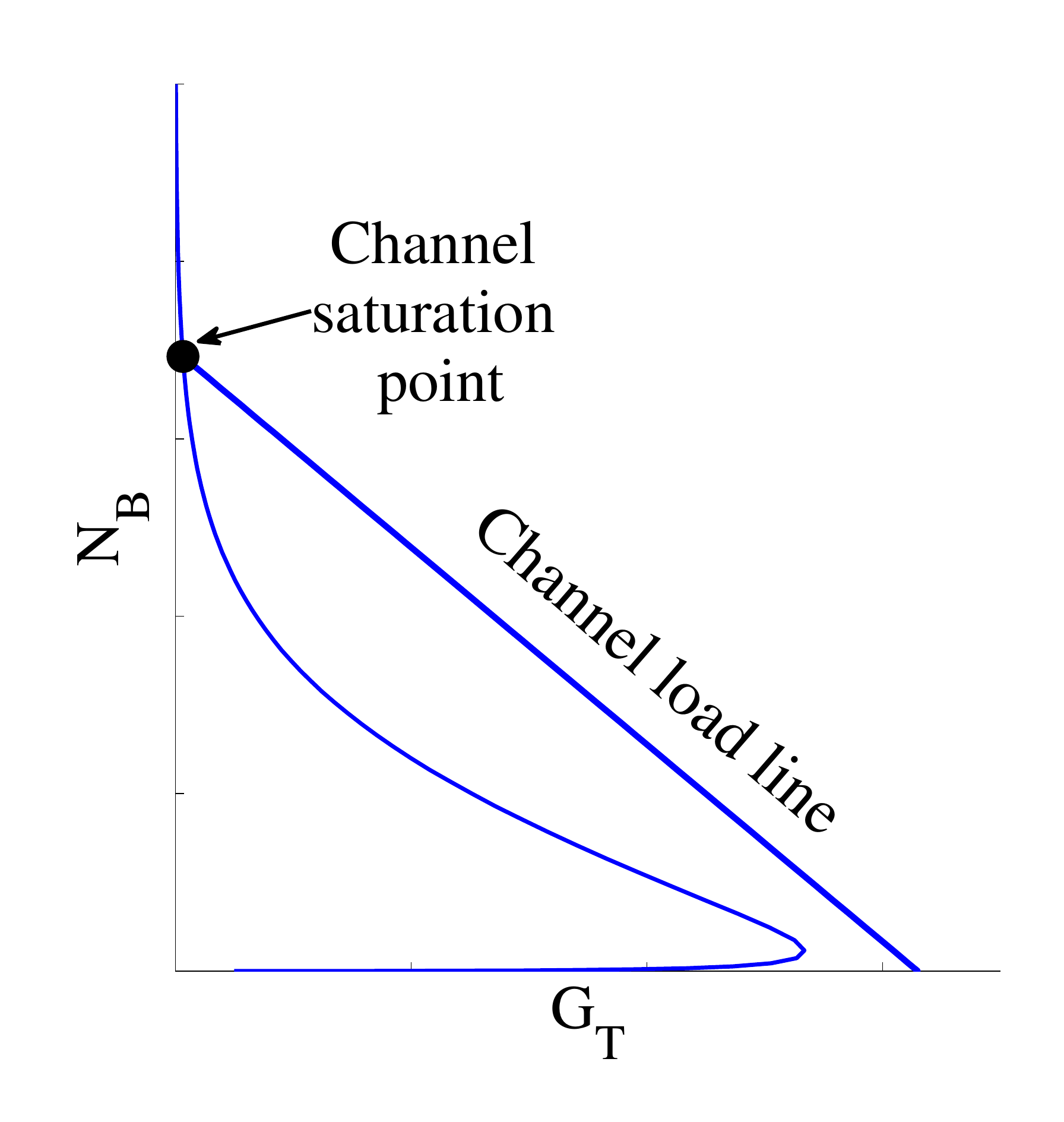}
\label{over}
} \qquad

\caption{\small{Examples of stable and unstable channels}}
\label{All_channels}
\end{figure}

Consider the examples in Figure~\ref{All_channels}. Each channel load line can intersect the equilibrium contour in one or more points (i.e. for one or more $N_B$ values). These intersections are referred to as equilibrium points since $G_{OUT}=G_T$ holds. The rest of the points of the channel load line belong to one of two sets: those on the left part of the plane with regard to the equilibrium contour represent points for which $G_{OUT}>G_T$, thus situations that yield to decrease of the backlogged population; those on the right part of the plane with regard to the equilibrium contour represent points for which $G_{OUT}<G_T$, thus situations that yield to growth of the backlogged population.  

Therefore, an intersection point is defined as a \textit{stable equilibrium point} with coordinates ($G_T^S$,$N_B^S$) if it enters the left part of the plane for increasing $N_B$ since for $N_B<N_B^S$ the result is that $G_{OUT}<G_T$ and for $N_B>N_B^S$ we find that $G_{OUT}>G_T$ so that the equilibrium point acts as a sink. With the same reasoning, an intersection point is defined as an \textit{unstable equilibrium point} with coordinates ($G_T^U$,$N_B^U$) if it enters the right part of the plane for increasing $N_B$. In this case it can be proven that as soon as a statistical fluctuation from the equilibrium point occurs, the number of backlogged users $N_B$ diverges from ($G_T^U$,$N_B^U$). As a matter of fact, as explained in \cite{CRDSA_stab1}, the model is based on the expected behavior while in reality the obtained values oscillate around the expected value.

In Figure~\ref{st} an example of stable equilibrium point is given. Being this point the only one of intersection it is also a \textit{globally stable equilibrium point} and we consider the related channel as stable since the communication will keep operating indefinitely around that point. On the other hand, if the point of equilibrium is not the only one as in Figure~\ref{unstFin} it takes the name of \textit{locally stable equilibrium point}. In particular the illustrated example shows two locally stable equilibrium points: one for a good throughput value, thus called channel operating point in the sense that is the point in which we want the communication to operate; one for throughput close to zero called channel saturation point since in that state too many users are in backlogged state and thus any packet transmission has hard times in being successful. Consider a scenario in which the communication starts from $N_B=0$. The communication will keep being around the operating point as long as statistical fluctuations are small enough to keep $N_B<N_B^U$. At a certain point however, the instability point will be crossed and in small time the saturation point will be reached. Depending on the communication settings, there is also a certain probability to exit from the state of saturation and come back around the channel operating point. However, this probability is generally small and considered negligible. Figure~\ref{unstInf} represents the same scenario as in Figure~\ref{unstFin} (i.e. the case of unstable channel) but for an infinite number of users. In this case the channel load due to new transmissions is independent on the actual number of backlogged users. Nevertheless the same discussion as for the case with finite number of users is valid and we can assume that the point of saturation is found for $N_B\rightarrow \infty$. Notice that in this case the formula for the channel load used so far is no longer valid. However, if Poisson arrivals with expected value $\lambda$ in terms of new packets to transmit are considered, the channel load line can be expressed as 
\begin{equation}\label{LL2}
G_T=\frac{\lambda \cdot \tau}{T_F}
\end{equation}
As a matter of fact, for a finite number of users the number of new transmissions is binomially distributed, while for a number of users that goes to infinity we can consider the binomial distribution converging towards the poissonian one.
Finally Figure~\ref{over} shows another example of globally stable equilibrium point. However, in this case the intersection point occurs for throughput close to zero. Therefore the point is defined as channel saturation point and the channel is considered overloaded.

\section{Packet Delay in stable channels}

Assuming a stable channel so that only a globally stable and operating point is present, it is of interest to know the delay associated to successfully transmitted packets. For a generic packet, it is possible to do so using the discrete-time Markov chain in Figure~\ref{user_state}. $T_F$ is assumed to be our discrete time unit. Therefore, the packet delay $D_{pkt}$ can be calculated as the number of frames that elapse from the beginning of the frame in which the packet was transmitted for the first time, till the end of the one in which the packet was correctly received.

\begin{equation}\label{FSMeq}
Pr\{D_{pkt}=f\} =
\begin{cases}
1-PLR\text{\,,\ \ \ \ \ \ \ \ \ \ \ \ \ \ \ \ \ \ for  } f=1 \\
\\
PLR\ [p_r\ (1-PLR)] \cdot \\
\cdot [1-p_r+PLR\ p_r]^{f-2}     \text{\ ,\ for  } f>1
\end{cases}
\end{equation}

Based on Equation~\ref{FSMeq} the expected packet delay can be written as

\begin{equation}\label{FSMAvEq}
Av[D_{pkt}]=\sum_{f=1}^{\infty}  f \cdot Pr\{D_{pkt}=f\}
\end{equation}

Equation~\ref{FSMAvEq} can also be rewritten in a form that is more practical for our analysis, by means of Little's Theorem. As a matter of fact, in a stable system the average number of users in B state is equal to the average time spent in backlogged state multiplied by the arrival rate of new packets $G_T$ (that we know to be equal to $G_{OUT}$ at the operational point). Therefore

\begin{equation}\label{LittleFor}
Av[D_{pkt}]=\frac{N_B^O\cdot \tau}{G_{OUT}^O\cdot T_F}
\end{equation}

where the presence of $T_F / \tau$ in the formula has the aim of normalizing the delay to the frame unit.

\section{Comparison of Random Access techniques}

Before starting the analysis of the results, it is useful to have a more solid comprehension of the role of three key parameters for the communication: the probability of new transmission $p_0$, the probability of retransmission $p_r$ and the total population $N_{TOT}$. The first two parameters have influence on the channel load line while the retransmission probability influences the shape of the equilibrium contour. In particular, defining a generic line $y=m\cdot x+q$ with $x=G_T$ and $y=N_B$, $p_0$ is inversely proportional to $m$. Therefore, fixing $q$, a decrement of $p_0$ determines a change for the slope of the channel load line that becomes steeper while an increment of $p_0$ has the opposite effect on the slope. $N_{TOT}$ has the same graphical meaning of $q$. In other words, fixing $p_0$ (i.e. the line slope) changing $N_{TOT}$ corresponds to changing the point of intersection with the y-axis since for $G_T=0$, $N_B=N_{TOT}$. Finally, as shown in Figure~\ref{changingPr}, a decrement of the retransmission probability determines a shift upwards of the equilibrium contour.     

\begin{figure}[tbh!]
\subfloat [Changing $p_0$] {\label{changingP0} \includegraphics [ scale = 0.23 ]{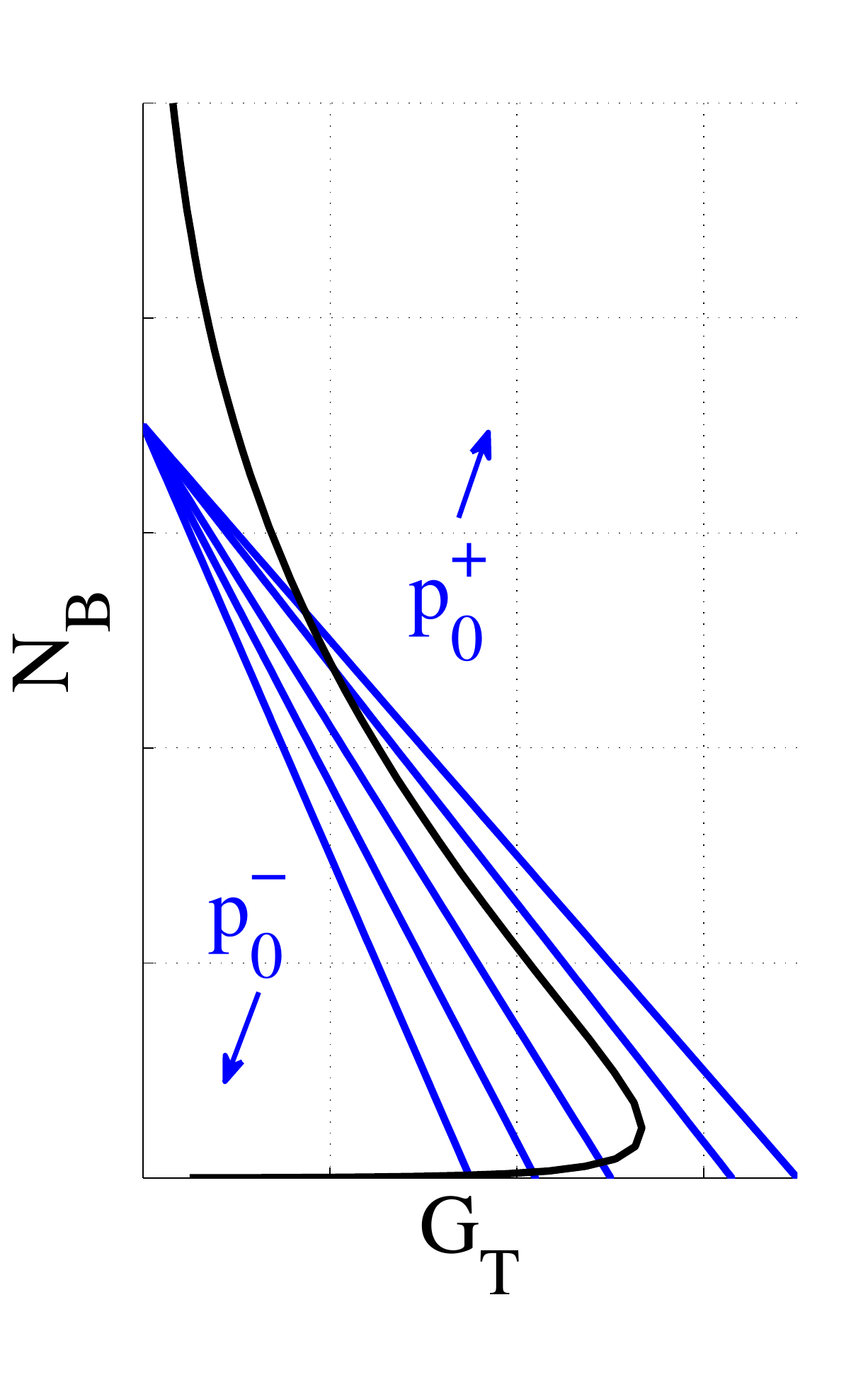}
\label{changingP0}
}
\subfloat [Changing $M$] {\label{changingM} \includegraphics [ scale = 0.23 ]{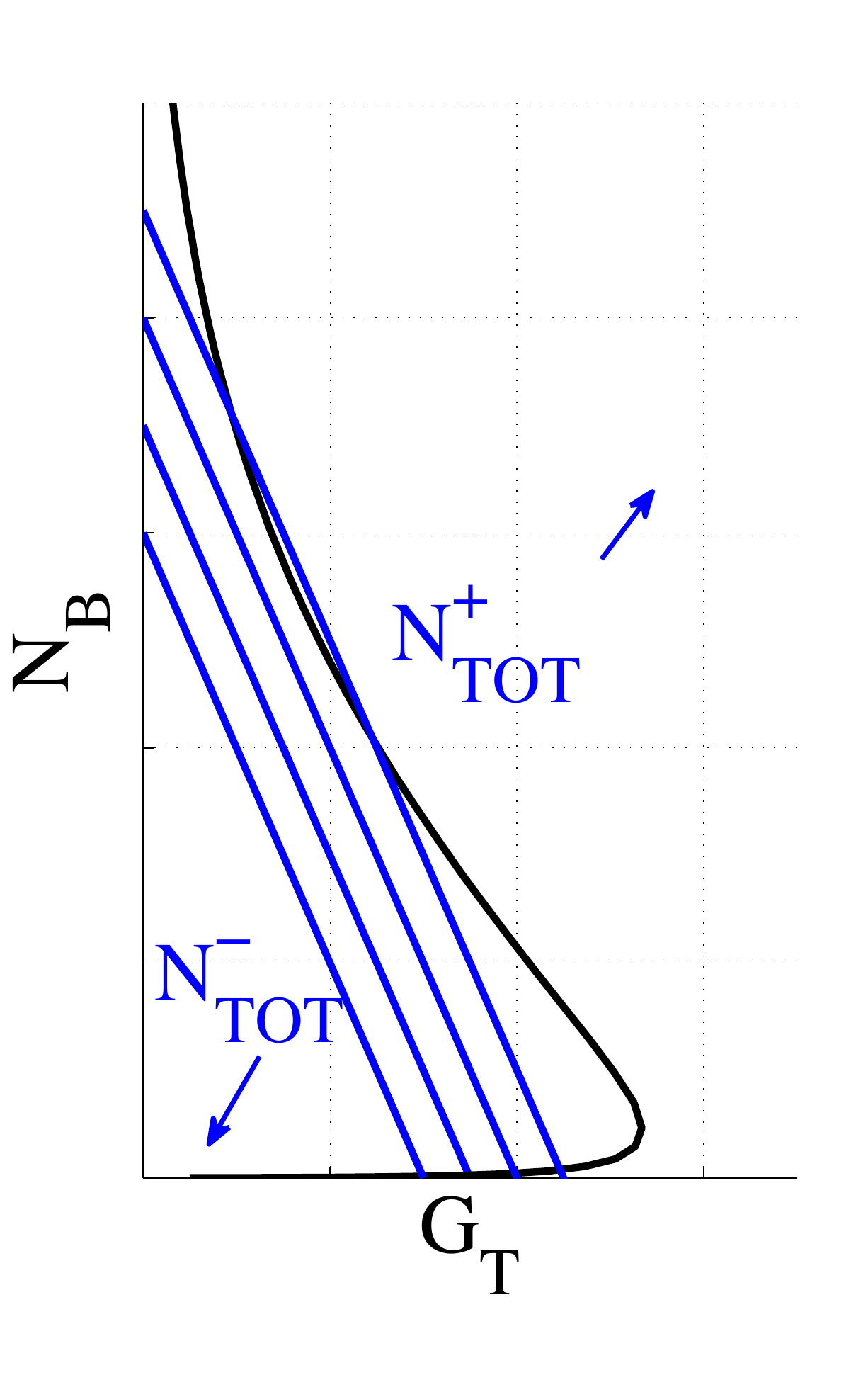}
\label{changingM}
}
\subfloat [Changing $p_r$] {\label{changingPr} \includegraphics [ scale = 0.23 ]{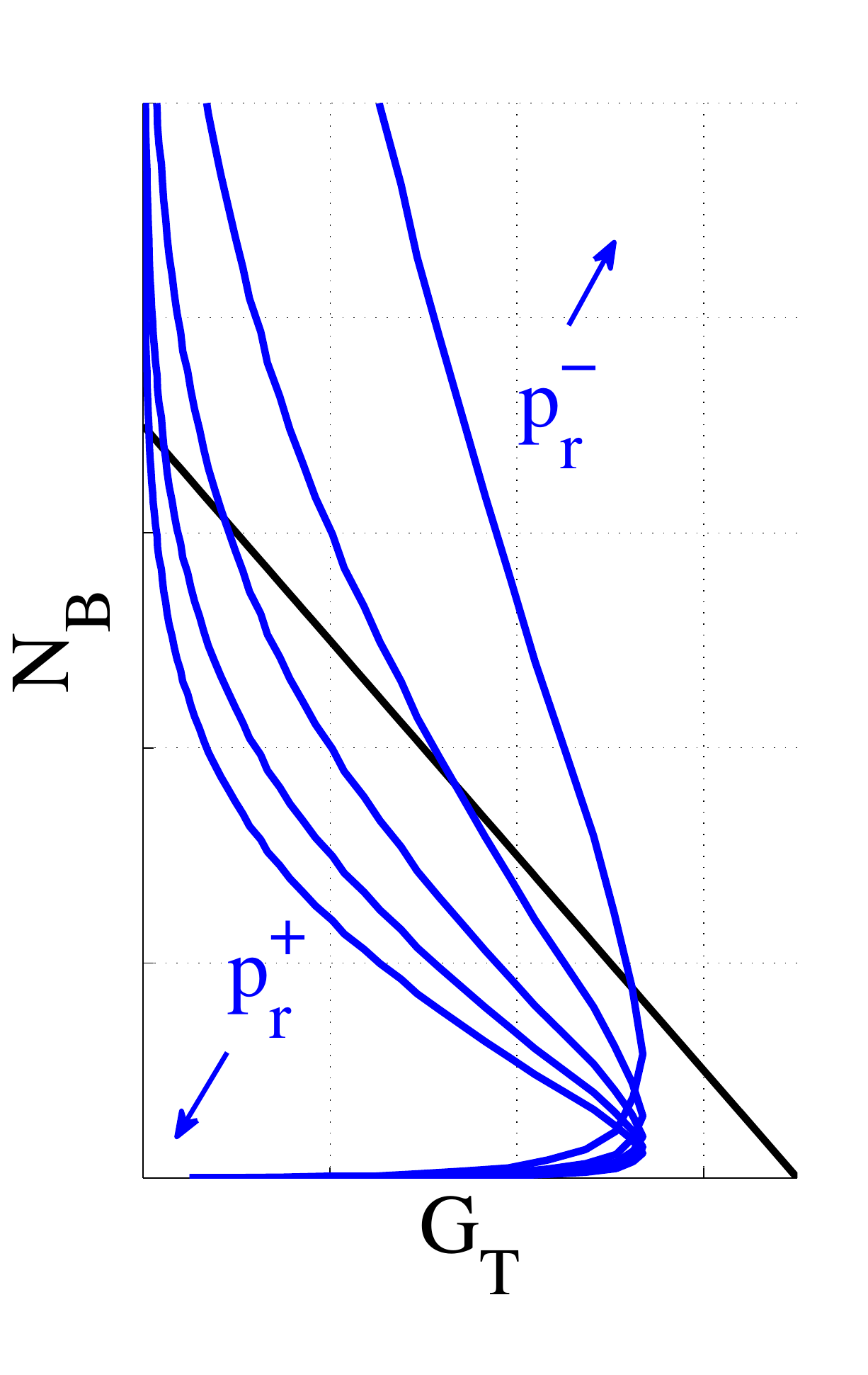}
\label{changingPr}
}
\\
\caption{\small{Graphical representation of the result for increments and decrements of $p_0$, $M$ and $p_r$}}
\label{All_changes}
\end{figure}

\begin{figure}[th!]
\centering
\includegraphics [width=1 \columnwidth] {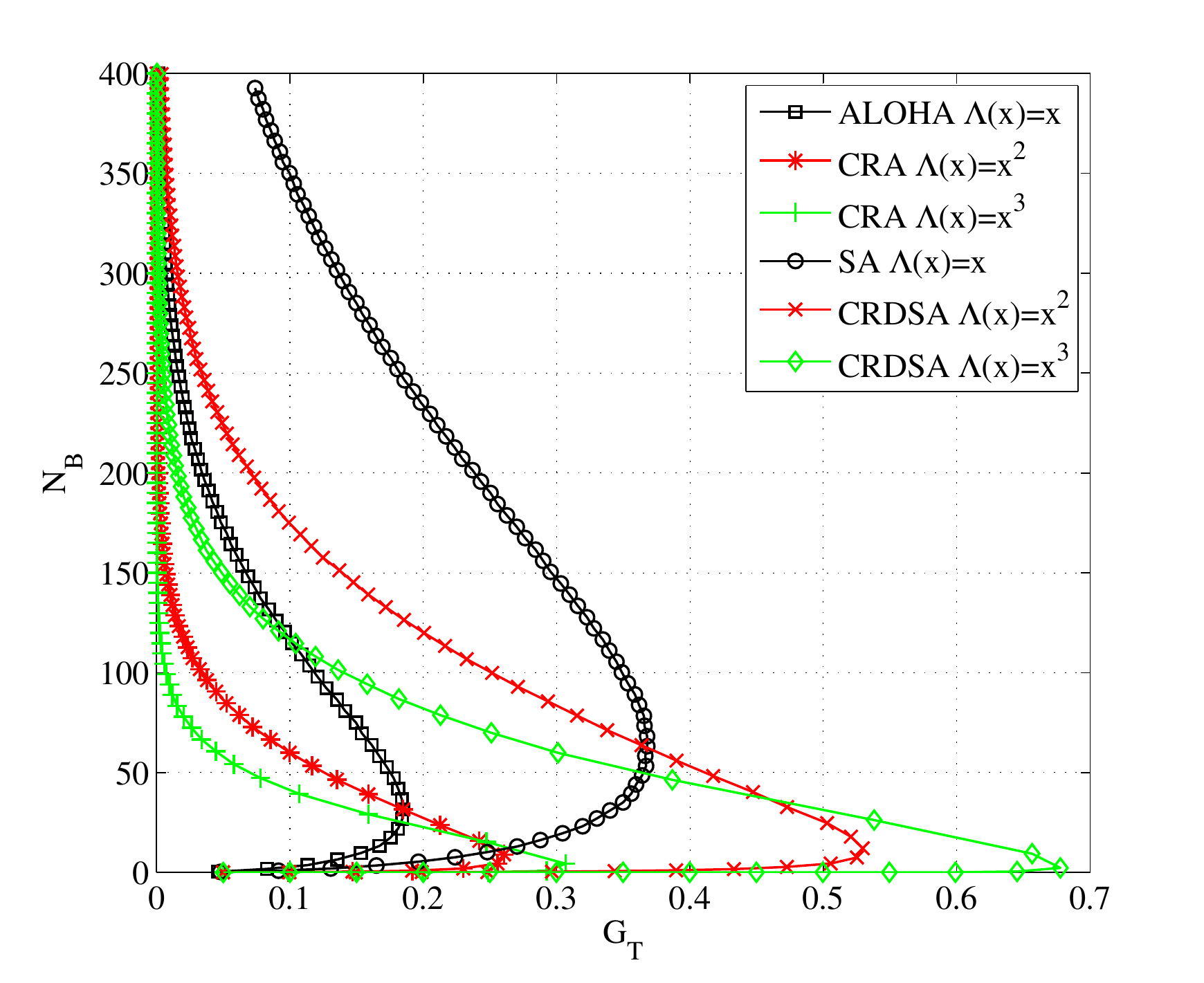}
\caption{\small{Equilibrium contour for pure ALOHA and CRA when no FEC is used.}}
\label{NoFEC}
\end{figure}

\begin{figure}[th!]
\centering
\includegraphics [width=1 \columnwidth] {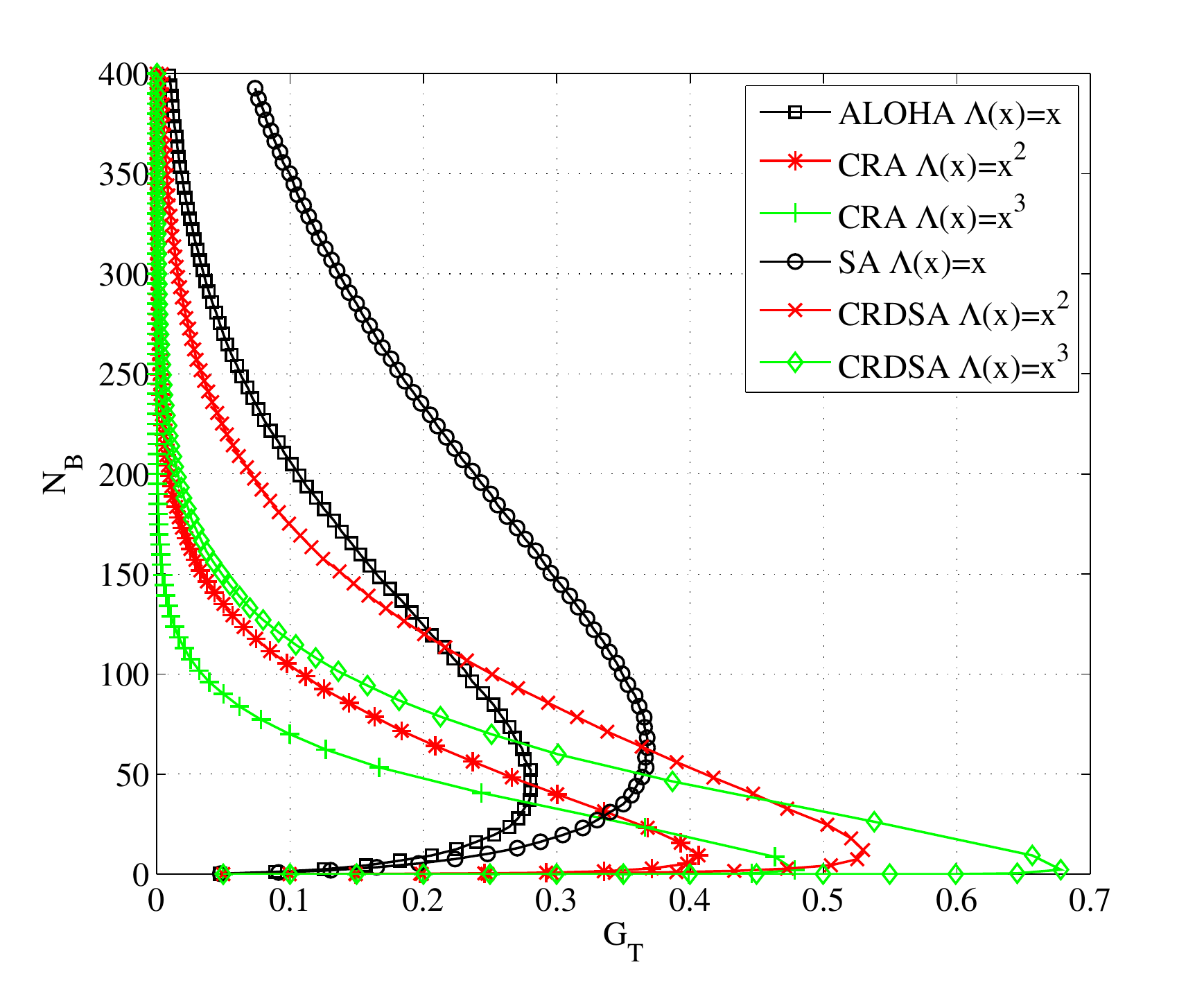}
\caption{\small{Equilibrium contour for ALOHA and CRA with associated FEC with $R_C=1/2$ and $SNR=2\ dB$}}
\label{2dB}
\end{figure}

\begin{figure}[th!]
\centering
\includegraphics [width=1 \columnwidth] {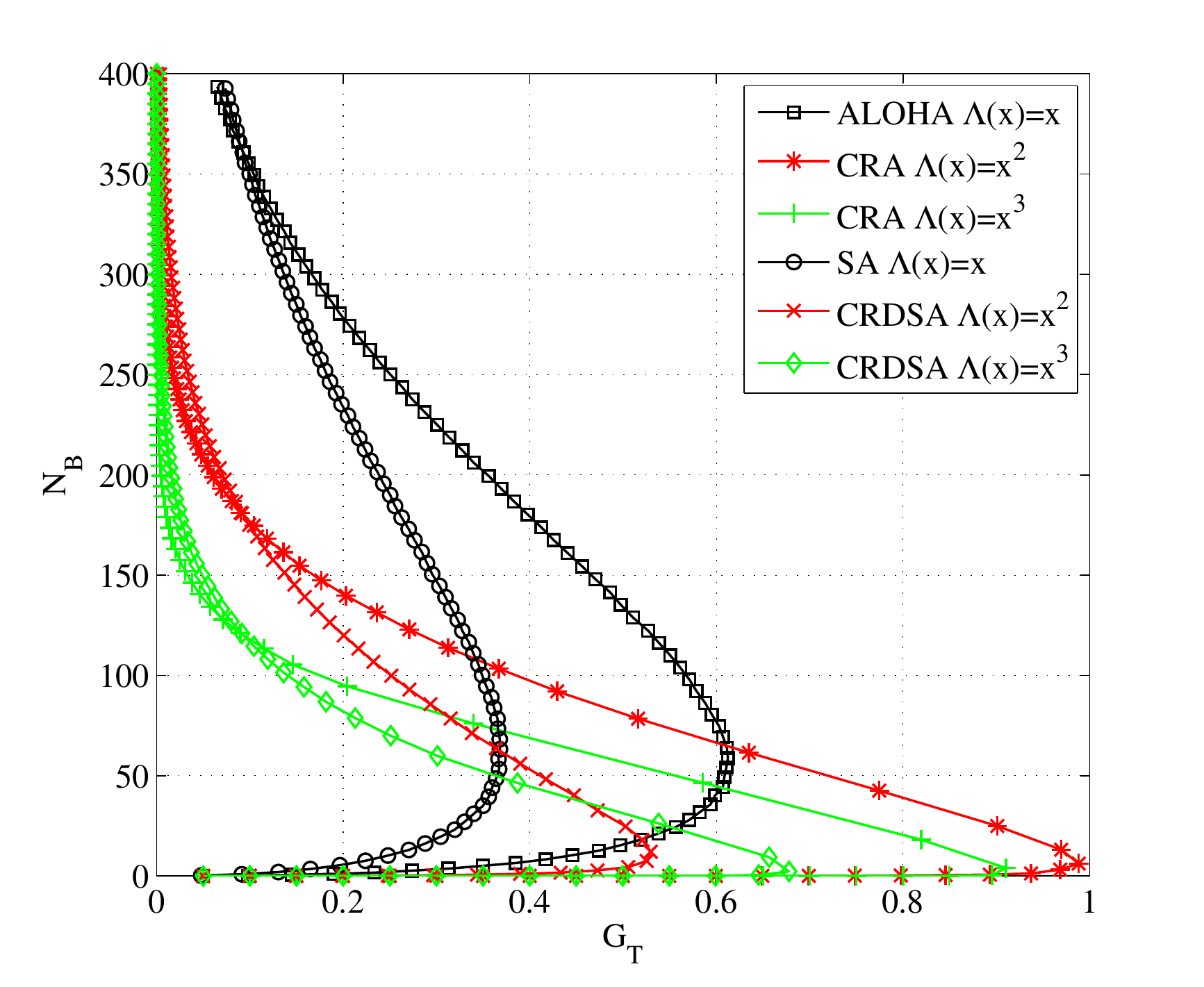}
\caption{\small{Equilibrium contour for ALOHA and CRA with associated FEC with $R_C=1/2$ and $SNR=10\ dB$.}}
\label{10dB}
\end{figure}

Figures~\ref{NoFEC}-\ref{10dB} show results for the settings outlined in Section~\ref{SO}. In the same figures, also the results for Slotted Aloha and CRDSA are reported, assuming the same settings and a comparable frame size of $100$ slots, since $\frac{T_F}{\tau}=100$. Notice that the aim of this section is to give a qualitative analysis rather than precise numerical results. In fact, the obtained results are based on the Shannon Bound while in practical implementations a real code must be considered. Therefore a quantitative analysis would be of unnoticeable importance. On the other hand a qualitative analysis is still of big value since it can prove the general validity of the technique and highlight pros and cons with regard to the state of the art. 

Figure~\ref{NoFEC} shows that even when no FEC is used, CRA can reach higher values of throughput than Pure Aloha, if the communication is designed properly so that the channel is stable. However, throughput results are far from those obtained for CRDSA. In addition, having a stable channel in CRA assumes that the total number of users participating is small enough so that only one point of intersection is present. For example, in the case of CRA with $\Lambda(x)=x^3$, if we want an expected throughput close to the peak (i.e. $T\simeq 0.3$) the total number of users must not be bigger than $N_{TOT}\simeq 60$; on the other hand we can see that for Pure Aloha, almost $400$ users can take part in the communication still ensuring a channel operating point around the throughput peak. If the design constraints require the use of CRA together with a bigger number of users, we know from Figure~\ref{changingPr} that it is possible to decrease the retransmission probability for backlogged users $p_r$. Nevertheless the stability comes at the cost of increased packet delay. This can be qualitatively understood considering that decreasing $p_r$, the peak throughput remains the same while the corresponding number of backlogged users $N_B$ increases. Therefore from Little's theorem an increase in the average packet delay is expected.
Finally, it can be seen that without the use of FEC the results of Aloha and CRA are worse than those for SA and CRDSA. As a matter of fact the results for synchronous techniques give place to equilibrium contour with identical shapes but bigger in value of throughput as well as in width of the curve below the peak. 
Similar considerations can be done in Figure~\ref{2dB} for the case in which FEC is used and the $SNR$ is quite low ($2\ dB$). In fact, concerning asynchronous techniques the same reasoning as for the previous case applies. Moreover, concerning the comparison with synchronous techniques, we can see that SA and CRDSA still outperform asynchronous techniques even though the performance of the two gets closer.

Finally for high SNR ($10\ dB$) as in Figure~\ref{10dB}, asynchronous techniques outperform synchronous ones. In particular, it can be noticed that while for CRDSA the burst degree distribution $\Lambda(x)=x^3$ is always better than $\Lambda(x)=x^2$, in CRA when the SNR is high enough $\Lambda(x)=x^2$ appears to be the best solution. However also in this case Pure Aloha still allows the participation in a stable communication of a higher number of users $N_{TOT}$ with regard to CRA.

\section{Conclusions}

In this paper a qualitative analysis of the stability in asynchronous Random Access schemes has been presented. In particular, stability results for CRA have been shown using a model based on the equilibrium contour. The obtained results have also been compared to Pure Aloha and CRDSA, showing that under the constraint of channel stability, despite the obtained throughput boost CRA supports a smaller number of users than pure ALOHA and does not appear convenient in low SNR scenarios with respect to synchronous access schemes. As a matter of fact, designing CRA to support a bigger number of users requires a decrement of the retransmission probability that yields to an increase on the average packet delay. Therefore further studies could investigate if this increment of packet delay would still allow asynchronous access schemes to be more efficient than Pure Aloha or not from a packet delay perspective.
We want to remark that obtained results for CRA represent an upper bound, since the Shannon Bound has been considered as decoding threshold for received bursts. This is the reason why in this paper the analysis has been accomplished in a qualitative and graphical manner rather than comparing the various techniques and burst degree distributions with numerical strictness. A very recent work proposed in \cite{ECRA} and called ECRA (Enhanced CRA) shows the possibility to outperform CRA in terms of throughput and Packet Error Rate and sets the more realistic Random Coding Bound as decoding threshold. While those results still do not constitute a practical case using a real code, they constitute an interesting step forward towards the case of a real scenario. The presented analysis can be as well extended to this recent evolution and future works should consider these latest findings rather than CRA.

\end{document}